\documentclass[authoryear]{elsarticle}
\usepackage[framed,autolinebreaks,useliterate]{mcode}
\usepackage{natbib}
\usepackage{amsmath}
\usepackage{rotating}

\usepackage{hyperref}
\usepackage{amssymb}
\usepackage{amsfonts}

\usepackage{pdflscape}
\journal{ }
\usepackage{color,transparent}
\usepackage{subfig}
\usepackage{caption}
\usepackage{setspace}
\usepackage{multirow}
\usepackage{array}
\newcommand\MyBox[2]{
  \fbox{\lower0.75cm
    \vbox to 1.7cm{\vfil
      \hbox to 1.7cm{\hfil\parbox{1.4cm}{#1\\#2}\hfil}
      \vfil}%
  }%
}

\begin{document}

% Title must be 150 words or less
\begin{frontmatter}
\title{A brain signature highly predictive of future progression to Alzheimer's dementia}

\author[a,b]{Christian~Dansereau\corref{cor1}}
\author[c]{Angela~Tam}
\author[a]{AmanPreet~Badhwar}
\author[c]{Sebastian~Urchs}
\author[a,e,f]{Pierre~Orban}
\author[d]{Pedro~Rosa-Neto}
\author[a,b]{Pierre~Bellec\corref{cor1}}
\author{for the Alzheimer's Disease Neuroimaging Initiative\corref{cor2}}

\cortext[cor1]{Corresponding authors: christian.dansereau@criugm.qc.ca, pierre.bellec@criugm.qc.ca}
\cortext[cor2]{Data used in preparation of this article were obtained from the Alzheimer's Disease Neuroimaging Initiative (ADNI) database (adni.loni.usc.edu). As such, the investigators within the ADNI contributed to the design and implementation of ADNI and/or provided data but did not participate in analysis or writing of this report. A complete listing of ADNI investigators can be found at: \url{http://adni.loni.usc.edu/wp-content/uploads/how_to_apply/ADNI_Acknowledgement_List.pdf}}

\address[a]{Centre de Recherche de l'Institut Universitaire de G\'eriatrie de Montr\'eal, Montr\'eal, CA}
\address[b]{D\'epartement d'Informatique et de recherche op\'erationnelle, Universit\'e de Montr\'eal, Montr\'eal,CA}
\address[c]{Integrated Program in Neuroscience, McGill University, Montr\'eal,CA}
\address[d]{Douglas Mental Health institute, McGill University, Montr\'eal,CA}
\address[e]{Centre de Recherche de l'Institut Universitaire en Sant\'e Mentale de Montr\'eal, Montr\'eal, CA}
\address[f]{D\'epartement de Psychiatrie, Universit\'e de Montr\'eal, Montr\'eal, CA}

%

% abstract typically 150 words
\begin{abstract}
Early prognosis of Alzheimer's dementia is hard. Mild cognitive impairment (MCI) typically precedes Alzheimer's dementia, yet only a fraction of MCI individuals will progress to dementia, even when screened using biomarkers. We propose here to identify a subset of individuals who share a common brain signature highly predictive of oncoming dementia. This signature was composed of brain atrophy and functional dysconnectivity and discovered using a machine learning model in patients suffering from dementia. The model recognized the same brain signature in MCI individuals, 90\% of which progressed to dementia within three years. This result is a marked improvement on the state-of-the-art in prognostic precision, while the brain signature still identified 47\% of all MCI progressors. We thus discovered a sizable MCI subpopulation which represents an excellent recruitment target for clinical trials at the prodromal stage of Alzheimer's disease.
\end{abstract}

%-- 
\end{frontmatter}

\section{Introduction}
Alzheimer's disease (AD) is the most common age-related neurodegenerative disorder. The typical progression of late-onset, sporadic AD comprises a lengthy preclinical stage, a prodromal stage of mild cognitive impairment (MCI), and a final stage of dementia. Usually, by the time patients suffer from dementia, severe and irreversible neurodegeneration has already occurred. In order to be effective, therapies should likely be initiated at earlier stages of the disease. For this reason, many works have aimed at finding biomarkers that can predict future progression to AD dementia at the prodromal or even preclinical stages \citep{Rathore2017review,Orban2017c}. Accurate prediction of progression from MCI to AD dementia has however proven to be challenging, likely due to the considerable heterogeneity in brain pathology underlying both of these conditions \citep{Rathore2017review}. We propose here to work around the heterogeneity issue by identifying a subset of individuals with MCI who share a homogeneous brain signature highly predictive of progression to AD dementia.

A clinical diagnosis of Alzheimer's dementia is primarily established on the basis of amnestic (e.g. memory) or nonamnestic (e.g. language, visual, executive) cognitive symptoms that interfere with the patient's activities of daily living. The diagnosis also requires the absence of evidence for concomitant neurological diseases that can substantially affect cognition, such as Lewy body dementia, fronto-temporal dementia or vascular dementia \citep{Mckhann2011}. MCI show a noticeable and measurable decline in cognitive abilities, including memory and thinking skills, yet this decline is not severe enough to qualify for dementia \citep{Petersen2014}. While MCI is considered an intermediate stage between the expected cognitive decline of normal aging and the more-serious decline of dementia, not all MCI patients progress to Alzheimer's dementia. Across 41 robust MCI cohort studies, an overall annual conversion rate of 6.5\% to Alzheimer's dementia was reported (Mitchell 2009). A modest conversion to dementia of 30-50\% even in long-term ($>5$ years) observational studies, highlights the heterogeneity present in the MCI population.

Imaging biomarkers and machine learning algorithms are increasingly used to complement neuropsychological testing for AD diagnosis and prognosis \citep{Dubois2007,Rathore2017review}. Established imaging biomarkers of AD are Positron Emission Tomography (PET) glucose metabolism, beta-amyloid and tau deposits \citep{Fodero2011,Sperling2011}, as well as non-invasive structural magnetic resonance imaging (sMRI) brain atrophy \citep{Lerch2005a}. Currently the state-of-the-art performance on the most popular reference dataset, assembled by the Alzheimer's disease neuroimaging initiative (ADNI), reaches 95\% accuracy to classify AD vs cognitively normal (CN) \citep{Fan2008a,Zhu2014,Xu2015,Zu2016}, and 80\% accuracy to identify patients with MCI who will progress to AD dementia in the next three years \citep{Mathotaarachchi2017,Moradi2015,Eskildsen2013,Wee2013,Gaser2013,Davatzikos2011,Koikkalainen2011,Misra2009}. Accuracy scores, however, are difficult to interpret in isolation. \cite{Korolev2016} for example, separately reported the specificity (76\%, proportion of stable MCI being correctly identified), sensitivity (83\%, proportion of progressor MCI being correctly identified), and precision (80\%, proportion of actual progressors amongst individuals identified as such). Precision is, in other words, the rate of progression in the subpopulation of MCI patients for which the machine learning algorithm makes a prognosis of dementia. Precision is thus a key metric for enrichment in clinical trials, as it dictates how many patients will decline in the absence of treatment. For a given sensitivity and specificity, the precision depends on the baseline rate of progression in the original MCI sample. The progression rate observed in the sample used in this paper (ADNI2) is 34\%, and corresponds with the range typically observed in other cohorts followed for over 3 years \citep{Mitchell2009}. Adjusted to a 34\% baseline progression, the precision levels reported so far in the literature ranged from 50\% to 75\%, see Table \ref{table_lit}. There is, therefore, substantial margin for improvements in terms of prognostic precision for AD dementia within 3 years, which is the focus of this work. 

The precision of imaging-based diagnosis of AD in past studies is likely limited by the pathophysiological heterogeneity of clinical diagnosis. The actual cause of dementia, AD or otherwise, can currently only be confirmed by a post mortem brain examination. The hallmarks of AD are the accumulation of beta-amyloid  plaques and tau protein neurofibrillary tangles in the brain, as well as marked atrophy of the medial temporal lobe. The analysis of \cite{Beach2012} revealed an important mismatch between clinical and histopathological diagnoses: sensitivity ranged from 71\% to 87\%, and specificity ranged from 44\% to 71\%, depending on the level of confidence in the clinical and pathophysiological examination. In particular, 30\% of patients diagnosed with AD dementia in that study had no or minimal signs of AD pathology, while markers of AD pathology has been observed in 10\% to 30\% of cognitively normal (CN) individuals, as well as 40\% of patients diagnosed with non-AD dementia \citep{Beach2012}. In the MCI population \cite{Petersen2014} reported prevalence of 4.8\% per year. In addition to such incorrect diagnoses, co-occurrence of other age-related neurodegenerative diseases is common, including vascular brain injury, Lewy body disease, or hippocampal sclerosis \cite{Rabinovici2017,Jellinger2014}. Individuals suffering from MCI in particular exhibit a wide range of brain pathologies \citep{Stephan2012}. In summary, the clinical diagnoses currently used are often incorrect (wrong underlying disease) and incomplete (missing several interacting diseases). Brain markers likely cannot be linked to clinical diagnoses with high precision in this context. 

In this work, we proposed a new machine learning model that worked around the issue of heterogeneity by identifying a subgroup of patients who (1) shared homogeneous brain abnormalities; and (2) had a highly predictable clinical diagnosis or prognosis. A cluster analysis was first used on structural and functional magnetic resonance images to identify subtypes of brain atrophy and functional connectivity in a sample mixing CN individuals with patients suffering from AD dementia. Using a novel two-step procedure, a model was trained to identify a brain signature mixing subtypes from different modalities, that was highly specific of patients with dementia. We then identified a subset of MCI patients presenting with this brain signature, and evaluated the rate of progression to dementia within 3 years in these individuals.

\section{Results}
\subsection*{Simple simulation}
\begin{figure*}[ht]%[htbp]
\begin{center}
\includegraphics[width=0.8\linewidth]{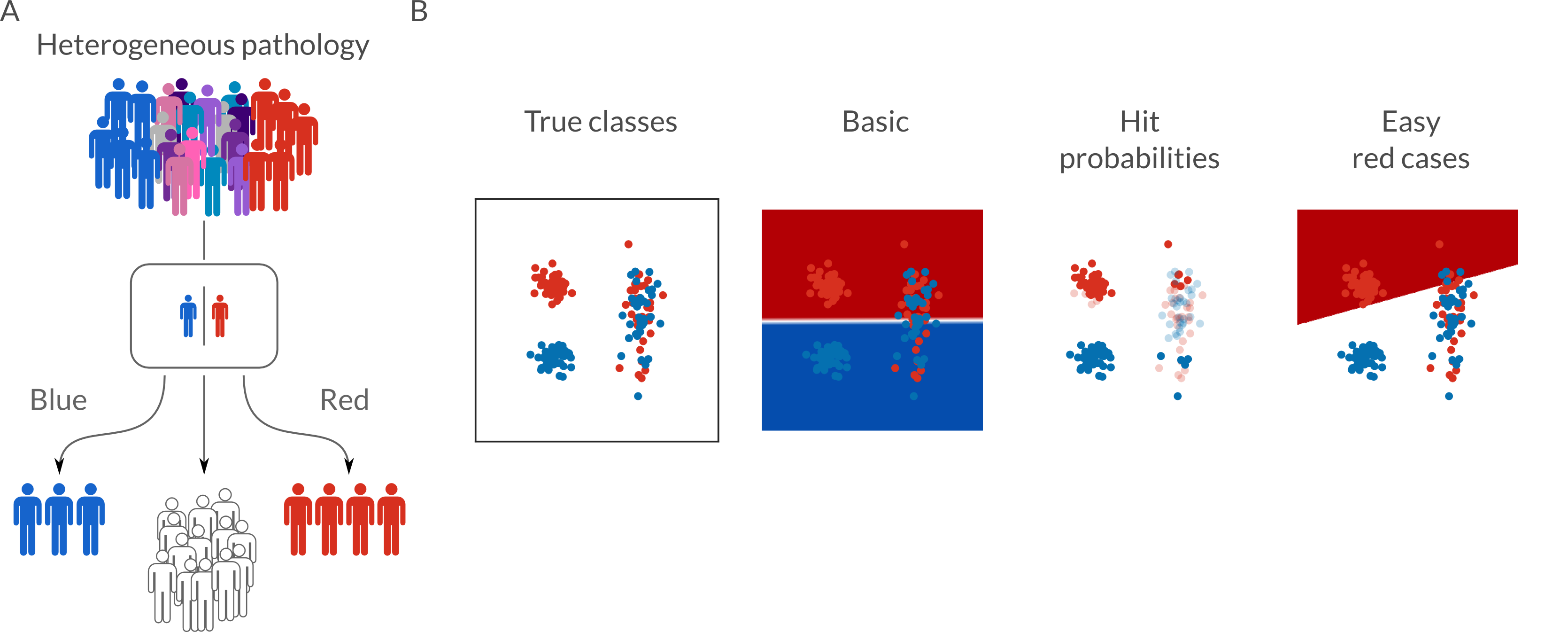}
\end{center}
\caption[HPS identification]{Panel A show the identification of easy cases for each class, Panel B prediction of clinical labels in a two-class problem, in the presence of heterogeneous labels in a subset of the data. The first column shows the initial classification problem with the distribution of the two classes. The second column shows a basic classifier decision hyperplane. The third column shows the subjects that have been flagged as high hit probability in hard color and the low hit probabilities with some transparency. The fourth column shows the final decision hyperplane of the red subjects with the HPS signature.
}

\label{fig_biotype_modes_toy}
\end{figure*}

We first illustrated the behaviour of the proposed method with a simple simulation (Figure \ref{fig_biotype_modes_toy}A). The task was to classify two classes using a separation line: blue dots for controls and red dots for patients. The distribution of both red and blue subjects was heterogeneous, in the sense that each distribution was a mixture of several Gaussian classes. Some of these classes were clearly separable, yet others were not, with blue and red points closely overlapping (maybe because of incorrect or incomplete diagnosis). When a standard classifier was applied on that data, it identified a separation line making a tradeoff in sensitivity and specificity across all examples (see Figure \ref{fig_biotype_modes_toy}B, second column). By perturbing the data, it was possible to identify the ``easy cases'', i.e. the data point that can be reliably classified correctly: more opaque points are associated with more reliable predictions and clearly identify the two well-separated classes at the top in Figure \ref{fig_biotype_modes_toy}B, third column. A separate model was then trained to identify the``easy cases'' red points (see Figure \ref{fig_biotype_modes_toy}B, fourth columns). The resulting prediction of red labels had limited sensitivity, as the problematic cases were not being detected at all, but it had near perfect specificity and precision. 
\subsection*{Multimodal imaging markers}
We extracted multimodal (structural and functional) measures of brain organization that could be used for automated AD diagnosis. The measures were derived from the baseline MRI scans of the ADNI2 cohort, which included anatomo-functional imaging for CN subjects (N=49) as well as patients suffering from AD dementia (N=24) (available sample size post quality-control on 10/2016). We decided to include a range of different measures previously shown to be sensitive markers of AD dementia. These included gray matter (GM) thickness \citep{Querbes2009,Eskildsen2013}, GM volume of various brain structures \citep{Karas2004}, as well as seed-based functional MRI (fMRI) connectivity maps generated for 20 intrinsic connectivity brain networks \citep{Urchs2017}.

\begin{figure*}[ht]
\centering
\includegraphics[width=\linewidth]{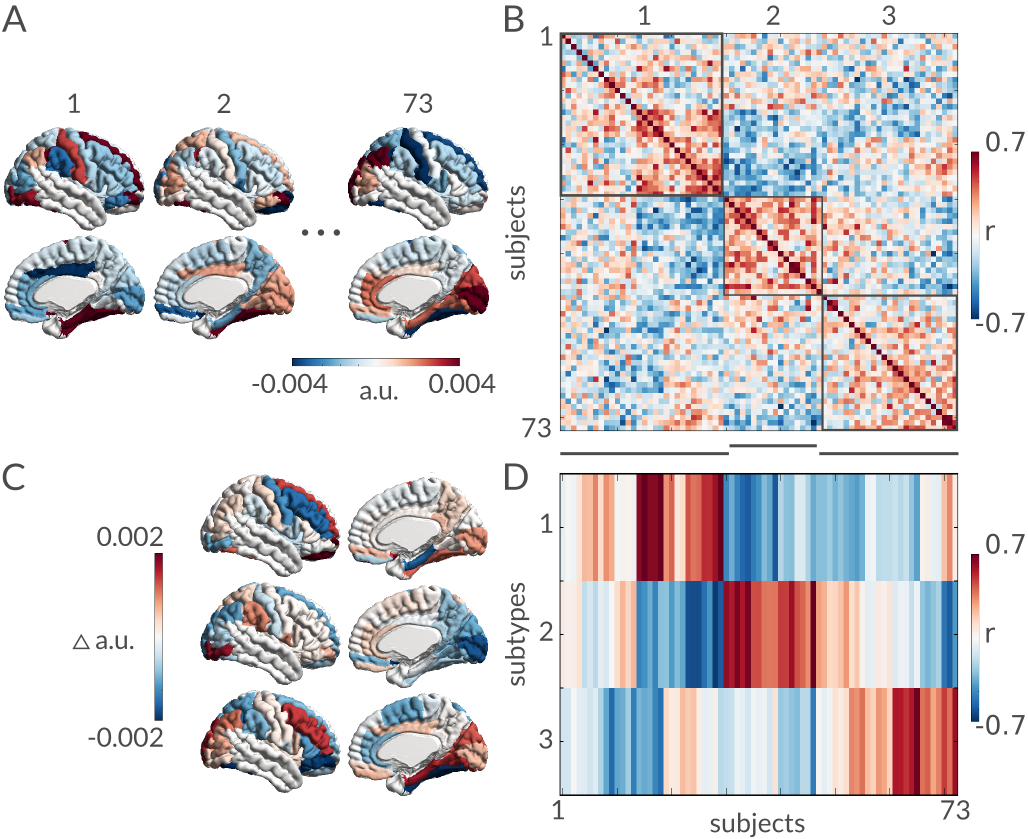}
\caption{Demeaned gray matter volume measures of the right hemisphere. Panel A shows individual maps and the correlation of every subject with all other subjects in Panel B. Panel C shows the subtypes templates representing subgroups in the dataset. Panel D shows the association of each individual map in A with each subtype template in C.}
\label{fig_subj_var}
\end{figure*}

Substantial inter-individual variations were observed in the distribution of normalized brain imaging measures. For example, some subjects showed higher- or lower-than average volumetric measures across extensive brain territories, such as the right medial occipital cortex in subject 1 (lower) and subject 73 (higher), see Figure \ref{fig_subj_var}A. We investigated whether such patterns could be found systematically in a subgroup of subjects. For this purpose, we quantified the similarity of GM volume maps between all pair of subjects using a Pearson correlation coefficient (Figure \ref{fig_subj_var}B). A cluster analysis revealed the presence of three subgroups of subjects with homogeneous GM volume maps. These subgroups were apparent as squares with high similarity values along the diagonal of the inter-subject similarity matrix, Figure \ref{fig_subj_var}B. These squares outline the spatial similarities of GM volume maps of subjects within a specific subgroup. By contrast, low similarity values were observed in elements outside of these squares, which corresponded to pairs of subjects falling into different subgroups. A subtype template was generated for each subgroup by averaging the maps of individuals within that subgroup, Figure \ref{fig_subj_var}B). In particular, subtypes 2 and 3 of GM volumetric maps reproduced the pattern observed in the occipital cortex of  subjects 1 and 73, respectively. The separation between clusters was not clear-cut in matrix \ref{fig_subj_var}B, suggesting a continuum rather than discrete subtypes. We thus extracted a continuous measure (Pearson's correlation) of similarity, called ``subtype weights'', between each individual map and each subtype map, Figure \ref{fig_subj_var}D). The subtyping procedure outlined above was applied independently for each type of measure (volumetric, cortical thickness, rs-fMRI) and each brain network (for rs-fMRI). We confirmed by visual inspection the presence of at least three subtypes for each modality/network, which we thus selected as a common number of subtypes across all modalities/networks for subsequent analyses.

\subsection*{Prediction of AD}
\begin{figure*}[ht]
\centering
\includegraphics[width=\linewidth]{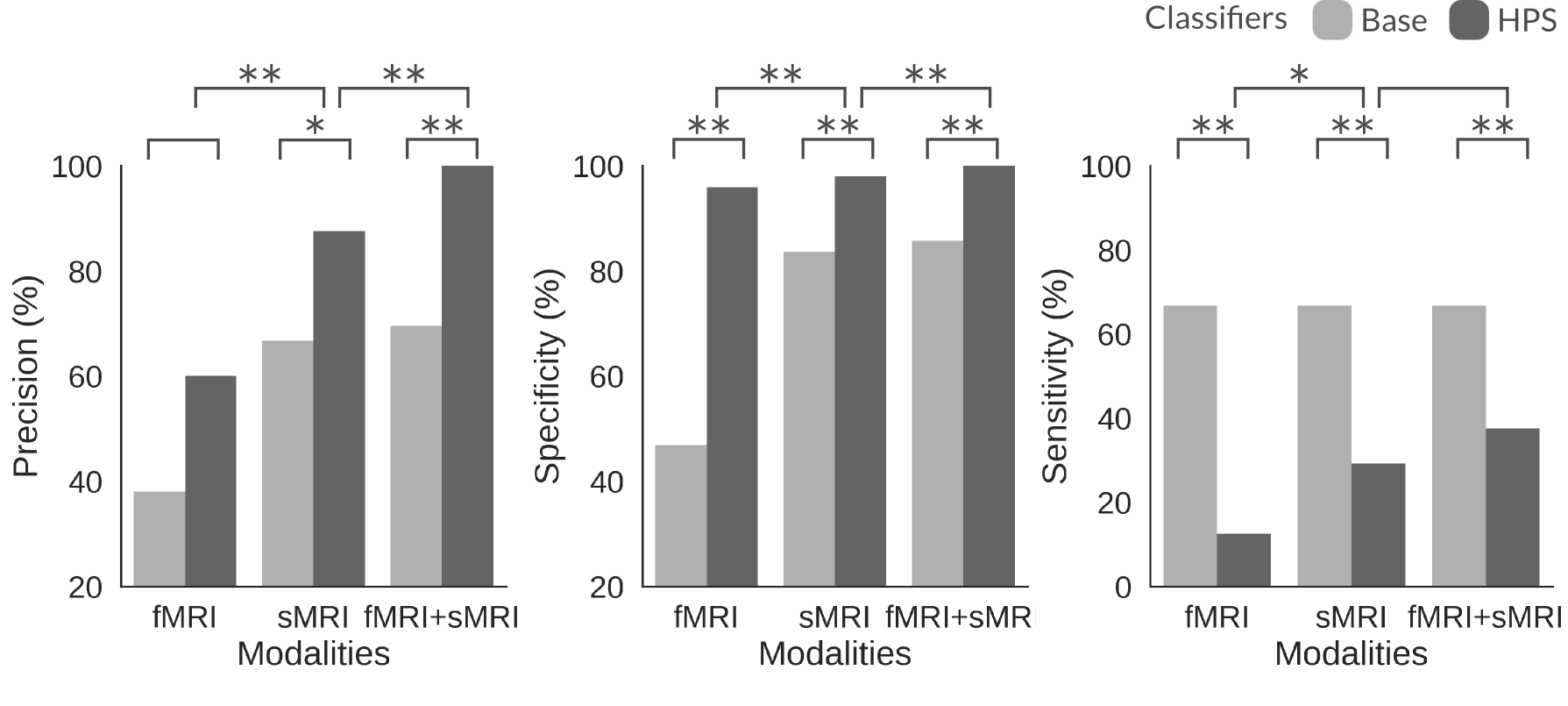}
\caption{Figure shows the precision, specificity and sensitivity of the three modalities (fMRI, sMRI and fMRI+sMRI) at each stage (Base: basic classifier and HPS: highly predictive signature). Significant differences are shown with $*$ for $p<0.05$ and $**$ for $p<0.001$).}
\label{fig_hpc_cnad}
\end{figure*}

We established a baseline performance for automatic classification of CN vs AD subjects using a well established machine learning model, i.e. a linear support vector machine model (SVM) \citep{Cortes1995}. The model reached 70\% precision (specificity 86\%, sensitivity 67\%) using tenfold cross-validation and multimodal (fMRI + sMRI) subtype weights, Figure \ref{fig_hpc_cnad}. Training only on fMRI subtypes or only on sMRI subtypes yielded lower performances: 38\% precision (specificity 47\% and sensitivity 67\%) for fMRI alone and 67\% precision (specificity 84\%, sensitivity 67\%) for sMRI alone. Note that, during cross-validation, the training of the model included both the generation of subtypes and the optimization of the SVM parameters.

\subsection*{Identifying easy cases}
As we outlined in the introduction, the core idea of our approach was to identify a subset of subjects for which clinical labels are easy to predict, such as the points on the left in Figure \ref{fig_biotype_modes_toy}A. To identify these ``easy cases'', we randomly perturbed the input data of the SVM model many times through subsampling, and assessed the hit probability for any given subject to be properly classified. We found that 68\% of individuals had a perfect (100\%) hit probability, with a small subset of subjects (18\%) exhibiting less reliable predictions (hit-probability $<90\%$), Supplementary material \ref{fig_hitproba}). We defined the ``easy cases'' as the subgroup of individuals reaching perfect hit probability. 
\subsection*{Predicting easy cases}
The next step of the method was to train a logistic regression \citep{Fan2008} to predict the AD ``easy cases'' Figure \ref{fig_methods}B, analogous to the rightmost column of Figure \ref{fig_biotype_modes_toy}B. The full multi-stage process of subtype extraction, hit probability estimation, and logistic regression was cross-validated using a ten-fold scheme in order to generate the performance of the prediction of AD ``easy cases''. A perfect 100\% precision (specificity 100\%, sensitivity 36\%) was reached for AD ``easy cases'', using multimodal structural and functional features. The multimodal HPS performance was a significant improvement (in precision and specificity, $p<0.001$) compared to the model trained on fMRI only, precision of 60\% (specificity 96\%, sensitivity 13\%), and sMRI only, precision of 88\% (specificity 98\%, sensitivity 29\%), see Figure \ref{fig_hpc_cnad}. Compared to the reference SVM model, with multimodal features, the precision of our proposed HPS model was improved by a wide margin (30\%, $p<0.001$),  as well as the specificity (15\%, $p<0.001$), at the cost of a marked loss in sensitivity (30\%, $p<0.001$). See Supplementary material Table \ref{tab_performance} for a list of the performance of each model.
\subsection*{Highly predictive brain signature}
\begin{figure*}[ht]
\centering
\includegraphics[width=\linewidth]{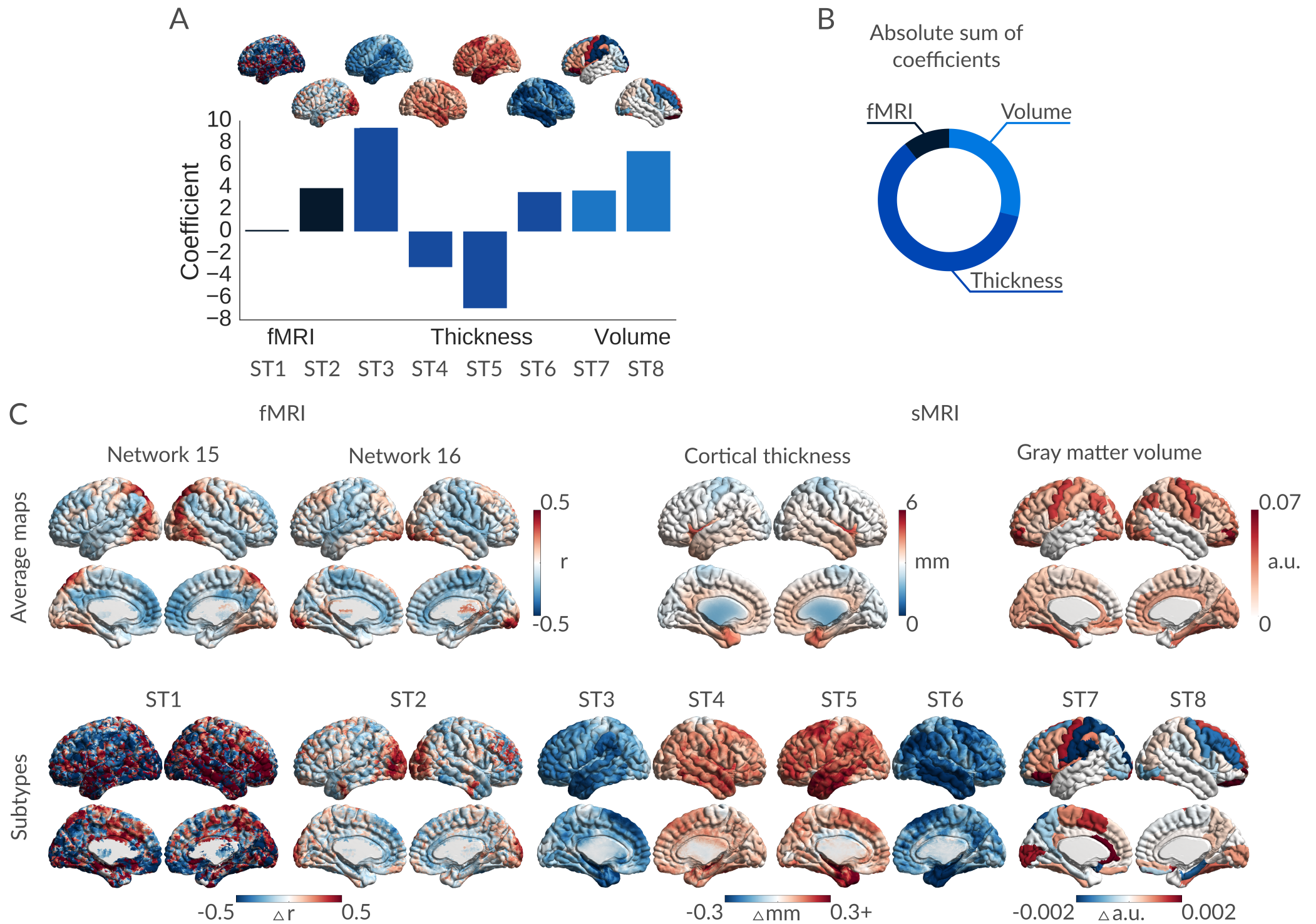}
\caption{Panel A shows the contribution of each modality to the decision, the ratios are computed by the sum of the absolute coefficient for each modality. Panel B shows the coefficients of the high-confidence prediction model for each subtype map. Panel C shows, on top, the average maps for each modality and on the bottom the subtype maps used for the high-confidence prediction.}
\label{fig_features_maps}
\end{figure*}

The logistic regression model used to predict AD ``easy cases'' is based on a set of coefficients, which give more or less weight to a particular subtype and modality. As such, the individuals flagged as AD ``easy cases'' can be seen as sharing a brain HPS, composed of a combination of subtype maps. The logistic model may in theory ignore a subtype or a modality entirely, by setting the corresponding weights to zero. In practice, we found that the HPS relied on all three types of measures (functional connectivity, cortical thickness, and gray matter volume), Figure \ref{fig_features_maps}A. To rank the contribution of each modality in the decision process, we computed the absolute sum of the coefficients for each measure, relative to the sum of all absolute coefficients (Figure \ref{fig_features_maps}B). The thickness was the most important measure (60\%), followed by the volumetric measures (29\%), and finally functional connectivity (11\%). The highest contributions came from four subtypes of thickness: bilateral patterns of cortical atrophy in temporal, sagittal and frontal areas (one subtype per hemisphere), and bilateral, opposite patterns of increased thickness (one subtype per hemisphere), Figure \ref{fig_features_maps}C. Two lateralized volumetric subtypes showed gray matter volume loss in the left motor, and right frontal areas as well as a gray volume increase in the left frontal and limbic regions. Finally, one functional subtype was very noisy and barely contributed to the model, while the other highlighted a connectivity subtype connecting the visual network with frontal areas. 
\subsection*{Prediction of progression to dementia}
\begin{figure*}[htbp]
\centering
\includegraphics[width=0.6\linewidth]{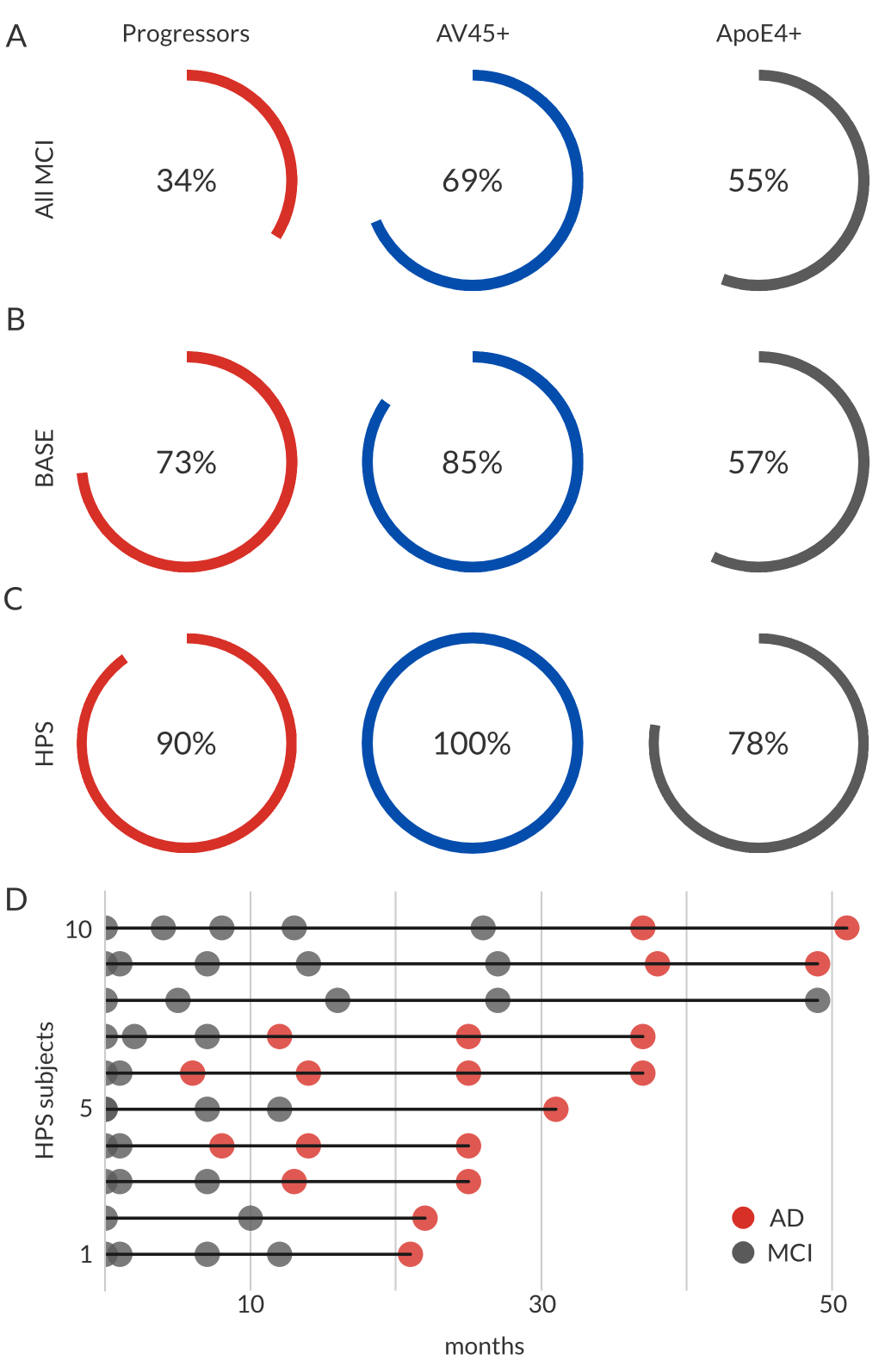}
\caption{Statistics on the MCI showing the signature. Panel A shows the percentage of MCI who progress to AD, the percentage of subjects positive for beta amyloid deposits using the AV45 marker and the percentage of carriers of one or two copies of the ApoE4 allele for the entire MCI cohort. Panel B shows the same statistics for the selection of the base classifier while Panel C displays statistics for subjects flagged as HPS. Panel D shows the clinical status of each HPS subject over time from the baseline scan.}
\label{fig_progressors}
\end{figure*}

We applied the HPS model to patients with MCI from the ADNI2 cohort, with the hypothesis that those with the signature would likely progress to AD dementia. The imaging sample for this experiment included the baseline structural and functional scans of all MCI patients in the ADNI2 cohort ($N=79$). We further stratified the patients with MCI into stable MCI (sMCI, $N=37$), i.e. most recent clinical status remains MCI with at least 36 months follow up, and progressors (pMCI, $N=19$), i.e. individuals whose most recent known clinical status is AD dementia, with progression from MCI to AD dementia occurring within 37 months. The HPS model selected a subset of 10 MCI subjects. Using the longitudinal follow-up clinical data provided by ADNI2, we found that 9 out of 10 of these subjects were pMCI (precision of 90\%, specificity of 97\%, sensitivity of 47\%), compared to 34\% pMCI in the whole MCI sample ($p<0.001$), Figure \ref{fig_progressors}A. Within the HPS subgroup, the time to progression from baseline to the first evaluation of AD dementia appeared uniformly distributed from 5 to 37 months, with 50\% subjects progressing after 24 months (Figure \ref{fig_progressors}C). In addition, 100\% of the MCI participants flagged as HPS were tested positive for beta amyloid deposition with AV45 testing, compared to a 69\% rate in the whole MCI sample ($p<0.05$), Figure \ref{fig_progressors}A. The rate of ApoE4 carriers in the HPS subsample was 78\%, compared to 55\% in the whole MCI group ($p>0.05$), Figure \ref{fig_progressors}A. A similar observation could be made regarding the rate of male of 70\% in the HPS subsample and 52\% in the whole MCI group ($p>0.05$). Finally the average age in the HPS group was of 74 years $\pm7$ and 71 years $\pm7$ for the whole MCI group ($p>0.05$).

\begin{table}[]
\centering
\caption{Supervised classification of MCI progression to AD dementia using the ADNI database. Progression time was establish if the the subject progresses to AD status in the next 36 months. Significant improvement of our method compared to each paper for the adjusted accuracy and precision (adjusted for a pMCI ratio of 34\% comparable to our sample) and specificity are shown with $*$ for $p<0.05$ and $**$ for $p<0.001$) and conversely significant decrease in sensitivity of our method compared to each paper. Adjusted accuracy (Acc), adjusted precision (Prec), specificity (Spec), sensitivity (Sens)}
\label{table_lit}
\begin{tabular}{llllllll}
                                       & N            & Acc      & Prec  & Spec   & Sens   \\
Article                                & sMCI/pMCI    &          & adjusted   &    &    \\ \hline
\textbf{Dansereau et al. (This paper)} & 37/19        & 80\%     & \textbf{90\%} & \textbf{97\%} & 47\%   \\
Mathotaarachchi et al. (2017)          & 230/43       & 82\%     & 74\%              & 87\%*           & 71\%*         \\
Korolev et al. (2016)                  & 120/139      & 79\%     & 65\%*            & 76\%**          & 83\%*       \\
Moradi et al. (2015)                   & 100/164      & 78\%     & 63\%*            & 74\%**          & 87\%**          \\ 
Eskildsen et al. (2013)                & 134/149      & 67\%**   & 52\%**           & 68\%**         & 66\%          \\ 
Wee et al. (2013)                      & 111/89       & 77\%     & 68\%*            & 84\%**         & 64\%          \\ 
Gaser et al. (2013)                    & 62/133       & 80\%     & 70\%*           & 84\%**         & 71\%*          \\ 
Davatzikos et al. (2011)               & 170/69       & 79\%     & 63\%*           & 71\%**        & \textbf{95\%}** \\ 
Koikkalainen et al. (2011)             & 215/154      & 73\%*    & 58\%*           & 71\%**        & 77\%*          \\ 
Misra et al. (2009)                    & 76/27        & 67\%**   & 51\%**         & 60\%**          & 80\%*          \\ 
\end{tabular}
\end{table}

\section{Discussion}

The main goal of this work was to develop an imaging-based AD diagnosis and prognosis with high precision and specificity. The proposed HPS approach did reach excellent performance in these respects, with 100\% precision and specificity when distinguishing patients with AD dementia from CN participants and 90\% precision, 98\% specificity when predicting which MCI patients would progress to dementia, up to three years before onset (see Table \ref{tab_performance}). These results represent a sizable and significant improvement in precision and specificity over previous models on this task, see Table \ref{table_lit}. The high specificity of the HPS model came at the cost of a limited sensitivity, which is significantly less than most recent published models, see Table \ref{table_lit}. Previous models from the literature could likely be tuned to work in a regime of high specificity and precision as well as the one proposed here. Actually, the base SVM classifier can achieve a precision close to our two-stage model, when adjusting the decision threshold (see Supplementary Material \ref{fig_roc_ps}B for ROC curves). The main contribution of this work was to demonstrate using proper cross-validation that a model can be trained in this regime of high precision and specificity. The second contribution was to show that the subset of patients at high risk of progression can be identified using a limited set of easily interpretable ``subtypes'', and that was made possible by the two-stage model. The results of \citep{Beach2012} suggest that only about half of patients diagnosed with AD dementia have clear AD brain markers post-mortem. The HPS was identified in 38\% of patients with AD dementia and 47\% of progressor MCI patients, which is consistent with the idea that the HPS model is picking on a typical brain presentation of AD that is already present at the prodromal stage of the disease.

The anatomical features selected by the method were in line with recent subtyping works, e.g. \citep{Hwang2015}, showing predominant atrophy in the temporal lobe, as well as the temporo-parietal juncture, in particular. The functional maps were more difficult to interpret, and seemed to capture some noise pattern. They still made a significant improvement in the performance of the HPS model. Because of the regularization in the logistic regression used to build the HPS model, features coming from different modalities did compete to be selected in the model. If redundant features existed, the ones with largest predictive power were selected by the classifier. This may explain why the selected functional subtypes did not involve the regions showing atrophy in the structural subtypes. We hypothesized that the HPS inferred from the AD vs CN prediction would also be useful to predict if a subject at the prodromal stage (MCI) would progress to dementia. Our results did validate this logic, but alternative strategies may be investigated in the future, e.g. training a model directly on the progressor vs stable MCI. 

A limitation of the present study was a moderate sample size, with $N=56$ patients suffering from MCI. While the ADNI database is large, resting-state fMRI has only been added to the protocol in the later stages of the study, ADNI GO and ADNI2. In addition, fMRI was only acquired on a third of the participants, even after it was added to the protocol. Because of the early role of synaptic dysfunction in AD, and the potential ability of fMRI to capture such dysfunction, we wanted to build an anatomo-functional diagnostic tool. But this choice did limit the sample size of our study since the selected subjects needed to have imaging data of the two modalities and pass their respective quality control assessment. Even with a larger sample size, another limitation of the ADNI dataset is that it does not reflect the diversity of cases observed in real-life clinical practice. Participants were in particular screened to exclude vascular dysfunction, which is a common comorbidity in AD \citep{Gorelick2011}. Resources with more inclusive enrollment criteria will help to better assess the generalizability of a biomarker-based AD diagnosis. 

The most direct application of the HPS model is its use for population enrichment in pharmaceutical clinical trials on AD \citep{Woo2017,Mathotaarachchi2017}. By recruiting patients who would normally progress to AD dementia, such an enrichment would increase the effect size of the drug while reducing the sample size needed to demonstrate efficacy and therefore would also reduce the cost of the trial. The HPS brain signature is not shared among all the AD dementia population (making it a subtype), but is common enough to represent a substantial portion of participants of interest (about a third of AD dementia subjects and half of MCI progressors). An alternative enrichment strategy, more geared towards generalizability, would be to only exclude subjects that will very likely not progress to AD dementia. The HPS method thus brings us closer to precision medicine by proposing a middle ground between traditional clinical cohorts and an entirely individual medicine.

In this manuscript, we focused exclusively on two MRI modalities. Our rationale was that MRI is non-invasive and already widely used in patient care of elderly populations. Beta amyloid and tau PET imaging, by contrast, are more expensive and less available, while lumbar punctures are invasive. Nevertheless, as shown in our results the combination of multimodal factors may help to improve precision, specificity and sensitivity. Since the sensitivity of each modality to abnormality may vary across the disease stages it may be beneficial to combine them to obtain complementary information. It will therefore be important in the future to see if a combination of PET imaging, blood tests targeting specific markers, cognitive scores, genetic factors, lifestyle factors, or others can help create stronger or multiple HPS that would in effect increase the sensitivity of the model at earlier stages of the of Alzheimer's disease.

\section{Acknowledgments}

Data collection and sharing for this project was funded by the Alzheimer's Disease Neuroimaging Initiative (ADNI) (National Institutes of Health Grant U01 AG024904) and DOD ADNI (Department of Defense award number W81XWH-12-2-0012). ADNI is funded by the National Institute on Aging, the National Institute of Biomedical Imaging and Bioengineering, and through generous contributions from the following: Alzheimer's Association; Alzheimer's Drug Discovery Foundation; BioClinica, Inc.; Biogen Idec Inc.; Bristol-Myers Squibb Company; Eisai Inc.; Elan Pharmaceuticals, Inc.; Eli Lilly and Company; F. Hoffmann-La Roche Ltd and its affiliated company Genentech, Inc.; GE Healthcare; Innogenetics, N.V.; IXICO Ltd.; Janssen Alzheimer Immunotherapy Research \& Development, LLC.; Johnson \& Johnson Pharmaceutical Research \& Development LLC.; Medpace, Inc.; Merck \& Co., Inc.; Meso Scale Diagnostics, LLC.; NeuroRx Research; Novartis Pharmaceuticals Corporation; Pfizer Inc.; Piramal Imaging; Servier; Synarc Inc.; and Takeda Pharmaceutical Company. The Canadian Institutes of Health Research is providing funds to support ADNI clinical sites in Canada. Private sector contributions are facilitated by the Foundation for the National Institutes of Health \footnote{\url{www.fnih.org}}. The grantee organization is the Northern California Institute for Research and Education, and the study is coordinated by the Alzheimer's Disease Cooperative Study at the University of California, San Diego. ADNI data are disseminated by the Laboratory for Neuro Imaging at the University of Southern California.

The computational resources used to perform the data analysis were provided by Compute Canada\footnote{\url{https://computecanada.org/}}. This project was funded by NSERC grant number RN000028 and the Canadian Consortium on Neurodegeneration in Aging (CCNA), through a grant from the Canadian Institute of Health Research and funding from several partners including SANOFI-ADVENTIS R\&D. CD is supported by a salary award from the Lemaire foundation and Courtois foundation. PB is supported by a salary award from ``Fonds de recherche du Qu\'ebec -- Sant\'e'' and the Courtois foundation.

\section{Materials and methods}

\subsection*{Dataset}
All functional and structural data were obtained from the Alzheimer's Disease Neuroimaging Initiative 2 (ADNI2) sample, a longitudinal standardized acquisition including three populations: cognitively normal subjects ($N=49$, 46\% male, $74\pm7$ years of age), patients with mild cognitive impairment ($N=56$, 51\% male, $72\pm7.5$ years of age) and patients with dementia due to AD ($N=24$, 46\% male, $72\pm7$ years of age). All participants gave their written informed consent to participate in the ADNI2 study, which was approved by the local ethics committee of participating institutions across North America. The consent form included data sharing with collaborators as well as secondary analysis. The present secondary analysis of the ADNI2 sample was approved by the local ethics committee at CRIUGM, University of Montreal, QC, Canada. All resting-state fMRI and structural scans were acquired on 3T Philips scanners with 8 channel head coils. We performed analyses on the first usable scan (typically the baseline scan) from ADNI2.

The acquisition parameters were as follows: structural scan 170 slices, voxel size 1x1x1.2 mm3, matrix size 256x256, FOV 256 mm2, TR 6.8 s, TE 3.09 ms, FA 9 degrees. A functional scan of 7 min, 48 slices,  voxel size 3.3x3.3x3.3 mm3, matrix size 64x64, FOV 212 mm2, TR 3 s, TE 30 ms, FA 80 degrees, No. volumes 140. For detailed information on the acquisition, see www.adni-info.org.

\subsection*{Extraction of functional features}
Each fMRI dataset was corrected for slice timing; a rigid-body motion was then estimated for each time frame, both within and between runs, as well as between one fMRI run and the T1 scan for each subject \citep{Collins1994}. The T1 scan was itself non-linearly co-registered to the Montreal Neurological Institute (MNI) ICBM152 stereotaxic symmetric template \citep{Fonov2011}, using the CIVET pipeline \citep{Ad-Dab'bagh2006}. The rigid-body, fMRI-to-T1 and T1-to-stereotaxic transformations were all combined to resample the fMRI in MNI space at a 3 mm isotropic resolution. To minimize artifacts due to excessive motion, all time frames showing a frame displacement, as defined in \cite{Power2012}, greater than 0.5 mm were removed. An average residual frame displacement was also estimated after scrubbing for further group analyses. A minimum of 50 unscrubbed volumes per run was required for further analysis (13 subjects were rejected). The following nuisance covariates were regressed out from fMRI time series: slow time drifts (basis of discrete cosines with a 0.01 Hz highpass cut-off), average signals in conservative masks of the white matter and the lateral ventricles as well as the first principal components (accounting for 95\% variance) of the six rigid-body motion parameters and their squares \citep{Giove2009,Lund2006}. The fMRI volumes were finally spatially smoothed with a 6 mm isotropic Gaussian blurring kernel. Datasets were preprocessed and analyzed using the NeuroImaging Analysis Kit - NIAK - version 0.12.17 (http://niak.simexp-lab.org), under CentOS with Octave (http://gnu.octave.org) version 3.6.1 and the MINC toolkit (http://bic-mni.github.io/) version 0.3.18. Preprocessing of MRI data was executed in parallel on the Guillimin supercomputer (http://www.calculquebec.ca/en/resources/compute-servers/guillimin), using the pipeline system for Octave and Matlab - PSOM \citep{Bellec2012}. Seed-based fMRI connectivity maps were obtained using a functional brain template of 20 networks covering the entire brain \citep{Urchs2017}. The Pearson's correlation between the average time series of each network and every voxel of the brain was computed to derive one functional connectivity map per network.

\begin{figure*}%[htbp]
\centering
\includegraphics[width=\linewidth]{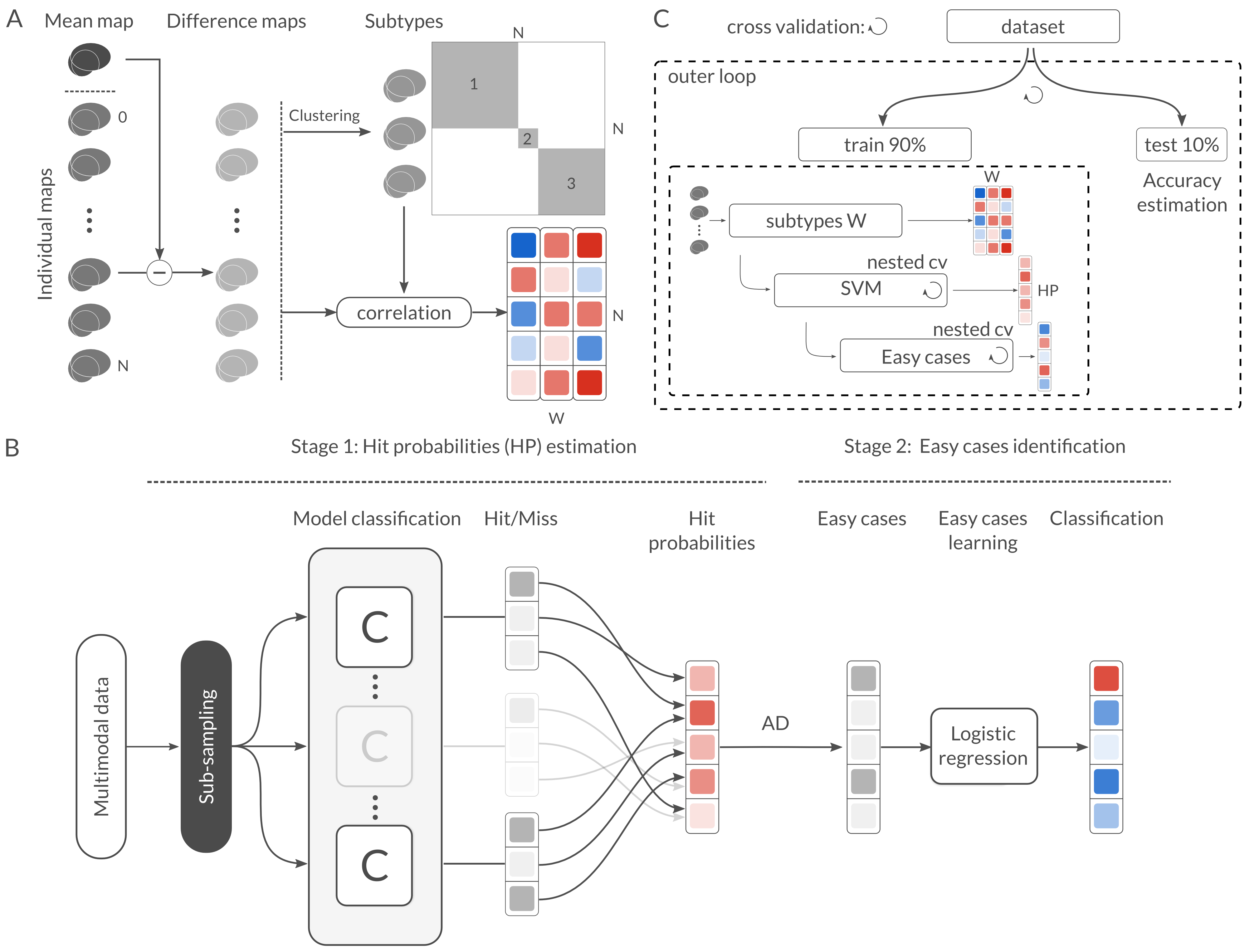}
\caption{Panel A shows the feature extraction method called subtypes weights, Panel B framework workflow: stage 1 shows the hit probability computation based on random sub-sampling and stage 2 shows the training of dedicated classifier for each “high-confidence” signature. Panel C shows the nested cross-validation scheme used in this method.}
\label{fig_methods}
\end{figure*}

\subsection*{Extraction of structural features}
Native individual T1-weighted MRI scans were corrected for non-uniformity artifacts with the N3 algorithm \citep{Sled1998}. The corrected volumes were then masked for brain tissues \citep{Smith2002} and registered into stereotaxic space \citep{Collins1994}. The registered images were segmented into gray matter (GM), white matter (WM), cerebrospinal fluid (CSF) and background using a neural net classifier \citep{Tohka2004}. The WM and GM surfaces were extracted using the Constrained Laplacian-based Automated Segmentation with Proximities algorithm \citep{Kim2005,Macdonald2000} and were resampled to a stereotaxic surface template to provide vertex based measures and lobar segmentation \citep{Lyttelton2007}. Cortical thickness was measured in native space using the linked distance between the two surfaces across 81,924 vertices \citep{Im2008}. Surface-based cortical thickness, as well as regional volume measures,  were obtained using the structural MRI images processed using the CIVET 1.1.12 pipeline for each hemisphere as described in \cite{Ad-Dabbagh2006}. The AAL template was applied on each hemisphere (40 regions per hemisphere) to extract the regional volumetric measures. The processing pipeline was executed on the Canadian Brain Imaging Network (CBRAIN) platform, a network of five imaging centers and eight High-Performance Computers for collaborative sharing and distributed processing of large MRI databases \citep{Frisoni2011}.

\subsection*{Multimodal imaging subtypes}
We extracted subtypes that characterize the interindividual variability within the sample comprising CN and AD participants (at the time of scanning), independently for each type of measure (functional maps, cortical thickness maps, and volumetric maps). In order to reduce the impact of some factors of no interest that may influence the clustering procedure, we regressed out the age, sex, and average post-scrubbing frame displacement from individual map, using a mass univariate linear regression model at each voxel. For each type of brain measure, we derived a spatial Pearson's correlation coefficient between all pairs of individual maps. This defined a subject x subject similarity matrix (of size 73 x 73), which was entered into a Ward hierarchical clustering procedure, as implemented in SciPy version 0.18.1 \citep{scipy,Walt2011numpy}. We selected three subgroups for each type of measure, based on a visual examination of the similarity matrix. For each type of measure, the average map of each subgroup defined a subtype. For each individual, we computed the spatial correlation of their map with each subtype. The resulting weight measures formed a matrix of size (number of subjects) x (number of subtypes), which was used as the feature space for all predictive models throughout the rest of this work. 
\subsection*{Prediction of AD}
The baseline prediction accuracy was obtained by training a SVM model with a linear kernel, as implemented in Scikit-learn \cite{scikit-learn} version 0.18. A tenfold cross-validation loop was used to estimate the performance of the trained model, including the entire subtyping procedure and regression of confounds. Classes were balanced inversely proportional to class frequencies in the input data for the training. A nested cross-validation loop was used (stratified shuffle split (50 splits, 20\% test size)) for the grid search of the hyper-parameter $C$ (grid was $10^{-2}$ to $10^{1}$ with 15 steps). Note that the $C$ parameter controlled how many misclassified examples the model will tolerate by adjusting the margin size. The model was evaluated using fMRI features only, sMRI features only, and the combination of fMRI and sMRI features. 
\subsection*{Identifying easy cases}
We randomly selected subsamples of the dataset (retaining 80\% of participants in each subsample) to replicate the SVM training 100 times. For each 80\% subsample, a separate SVM model was trained to predict the clinical labels (CN or AD), see Figure \ref{fig_methods}B. Note that the optimal $C$ parameter was estimated once using the whole available sample, as described above, and used across all subsamples. This was done to avoid creating major uncontrolled algorithmic variations. The linear discriminating weights of the SVM were still optimized independently for each subsample. Predictions of clinical labels were then made on the remaining 20\% of subjects, that were not used for training. For a given individual, the hit probability was calculated as the frequency of correct clinical classification across all available SVM replications where the test set included that individual. Easy cases were defined as individuals with 100\% hit probability.
\subsection*{Predicting easy cases}
We trained a logistic regression classifier \cite{Fan2008} to predict the AD easy cases. The logistic regression was trained using a L1 regularization on the coefficients, see Figure \ref{fig_methods}B. Class weight was balanced inversely proportional to class frequencies in the input data. A stratified shuffle split (100 splits, 20\% test size) was used to estimate the performance of the model for the grid search of the hyper-parameter C (grid was $10^{-0.2}$ to $10^{1}$ with 15 equal steps). In this case, the $C$ parameter controlled the sparseness of the weights.

\subsection*{Cross-validation}

A nested cross-validation was performed for accuracy estimation and parameters optimization. The outer loop used to estimate the generalizability of the framework was a ten-fold cross-validation scheme. Each training fold included the full multi-stage process of subtype extraction, SVM prediction of clinical labels, identification of HPS and prediction of HPS with logistic regression. Sensitivity (true positive rate, TP), specificity (true negative rate, TN) and precision $(TP/(TP+(1-TN)))$ of the diagnosis were estimated across all test folds, in the AD vs CN prediction. Cross-validation nested inside the outer loop was used to search for the optimal hyper-parameters, Figure \ref{fig_methods}C. 
\subsection*{Prediction of progression to dementia}
The HPS was obtained by applying the subtyping and easy cases recognition to the whole CN and AD sample, and considering all subtypes associated with non-zero weights by the sparse logistic regression model in Figure \ref{fig_methods}C stage 2. The logistic regression trained on AD vs CN was used to identify MCI patients who have a HPS of AD dementia. The imaging sample for this experiment included the baseline structural and functional scans of all patients with MCI in the ADNI2 cohort, with at least 36 months of follow-up ($N=56$). We further stratified the patients with MCI into stable MCI (sMCI, $N=37$)), i.e. latest clinical status is MCI, and progressors (pMCI, $N=19$), i.e. individuals whose most recent known clinical status is AD dementia, with progression from MCI to AD dementia occurring within 36 months. Note that no AV45 imaging data or genetic data, nor any data from the MCI cohort, were used to build the HPS model.
\subsection*{Statistical test of differences in model performance}
We generated a confidence interval on the performance (i.e. precision, specificity and sensitivity) of a given model using a Monte-Carlo simulation. Taking the observed sensitivity and specificity, and using similar sample size to our experiment, we replicated the number of true and false positive detection 100000 times using independent Bernoulli variables, and derived replications of precision, specificity and sensitivity. By comparing these replications to the sensitivity, specificity and precision observed in other models, we estimated a p-value for differences in model performance \citep{Phipson2010}. A p-value smaller than 0.05 was interpreted as evidence of a significant difference in performance,  and 0.001 as a strong evidence. This approach was first used in Figure \ref{fig_hpc_cnad} to contrast the performance of the HPS model to the baseline (SVM) model, both for AD vs CN and MCI progressor vs stable, as well as contrasting the performance of multimodal (fMRI+sMRI) model vs models using only fMRI or sMRI features. The same approach was used to contrast our proposed model for MCI progressor vs stable with results from the literature, in Table \ref{table_lit}. Note that, based on our hypotheses regarding the behaviour of the HPS model, the tests were one-sided for increase in specificity and precision, and one-sided for decrease in sensitivity. 
\subsection*{Statistical test of enrichment }
The HPS model was used to select a subset of the MCI population. We tested statistically if this subgroup was enriched for (1) progression to dementia; (2) AV45+, and; (3) ApoE4+. We implemented for this purpose a Monte-Carlo simulation, where we selected 100000 random subgroups out of the original MCI sample. By comparing the proportion of progressors (respectively AV45+ and ApoE4+) in these null replications to the actual observed values in the HPS subgroup, we estimated a p-value \citep{Phipson2010} (one sided for increase). A p-value smaller than $0.05$ was interpreted as evidence of a significant enrichment, and $0.001$ as a strong evidence.
\subsection*{Public code and data}
The code used in this experiment is available on a 
GitHub repository at the following URL\footnote{\url{https://github.com/cdansereau/HPS}}. An IPython Notebook is also provided with all of the figure generation scripts. Scikit-learn \cite{scikit-learn} version 0.18 was used for most of the machine learning algorithms and Nilearn \cite{Abraham2014} version 0.2.6 for visualization purposes.
\subsection*{ADNI dataset}
Data used in the preparation of this article were obtained from the Alzheimer's Disease Neuroimaging Initiative (ADNI) database (adni.loni.usc.edu). The ADNI was launched in 2003 as a public-private partnership, led by Principal Investigator Michael W. Weiner, MD. The primary goal of ADNI has been to test whether serial magnetic resonance imaging (MRI), positron emission tomography (PET), other biological markers, and clinical and neuropsychological assessment can be combined to measure the progression of mild cognitive impairment (MCI) and early Alzheimer's disease (AD).

\section*{References}

\bibliographystyle{elsarticle-harv}
\bibliography{biblio}

\pagebreak

\clearpage
\appendix

%% SUPPLEMENTARY MATERIAL
\clearpage
\pagebreak
\renewcommand{\thefigure}{S\arabic{figure}}
\renewcommand{\thetable}{S\arabic{table}}
\setcounter{figure}{0}
\begin{center}
\emph{Supplementary Material {--} A brain signature highly predictive of future progression to Alzheimer's dementia}\\
\end{center}

\begin{table}[htbp]
\centering
\caption{Performance of the models. Prec: precision, Spec: specificity, Sens: sensitivity and N: number of selected subjects.}
\label{tab_performance}
\begin{tabular}{lllllll}
Modality & Algo & Contrast & Prec (\%)& Spec (\%)& Sens (\%)& N \\ \hline
\multirow{1}{*}{\begin{tabular}[c]{@{}l@{}}fMRI\end{tabular}} & \multirow{1}{*}{Base} & CN/AD & 38.10 & 46.94 & 66.67 & 42 \\
 & \multirow{1}{*}{HPS} & & 60 & 95.92 & 12.5 & 5 \\
\multirow{1}{*}{\begin{tabular}[c]{@{}l@{}}sMRI\end{tabular}} & \multirow{1}{*}{Base} & & 66.67 & 83.67 & 66.67 & 24 \\
 & \multirow{1}{*}{HPS} & & 87.50 & 97.96 & 29.17 & 8 \\
\multirow{1}{*}{\begin{tabular}[c]{@{}l@{}}fMRI+sMRI\end{tabular}} & \multirow{1}{*}{Base} & & 69.57 & 85.71 & 66.67 & 23 \\
 & \multirow{1}{*}{HPS} & & 100 & 100 & 37.50 & 9 \\
\multirow{1}{*}{\begin{tabular}[c]{@{}l@{}}fMRI+sMRI\end{tabular}} & \multirow{1}{*}{Base}& sMCI/pMCI & 73.33 & 89.19 & 57.89 & 15 \\

 & \multirow{1}{*}{HPS}& & 90 & 97.3 & 47.37 & 10
\end{tabular}
\end{table}

\begin{figure*}%[ht]
\centering
\includegraphics[width=0.5\linewidth]{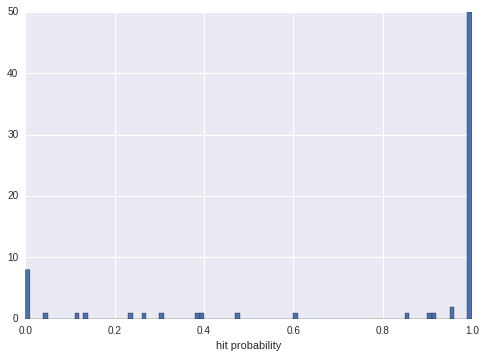}
\caption{Hit-probability distribution obtained from replicating the SVM training 100 times from 80\% of the training set.}
\label{fig_hitproba}
\end{figure*}

\begin{figure*}%[ht]
\centering
\includegraphics[width=\linewidth]{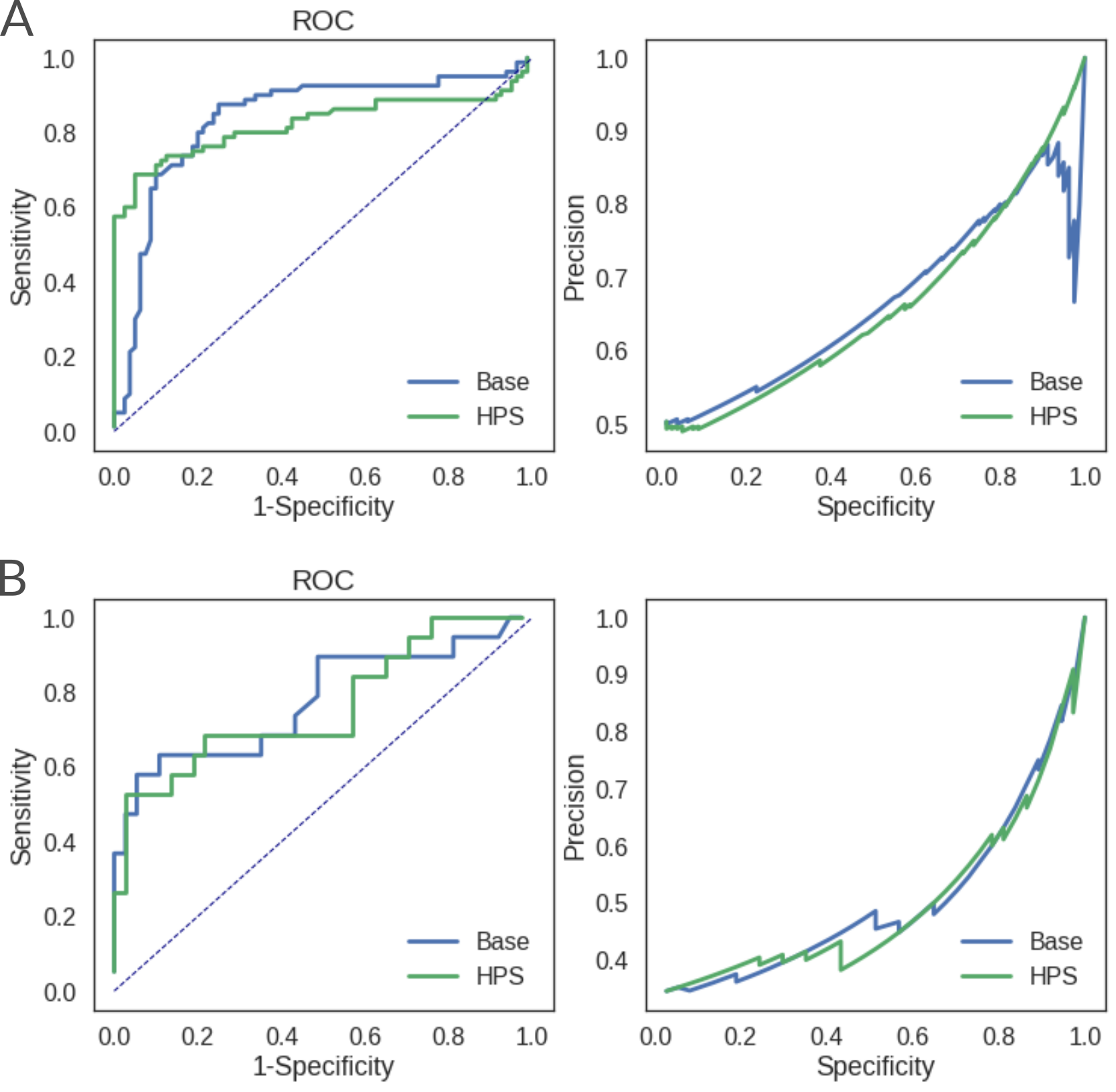}
\caption{Panel A show the ROC curve of the base classifier and HPS classifier for the simulation example of Figure \ref{fig_biotype_modes_toy} and on the right the precision vs specificity curves. Panel B show the same curves for the MCI progressors task.}
\label{fig_roc_ps}
\end{figure*}

\end{document}